\documentclass[11pt]{article}
\usepackage{amsmath,amsfonts}

\begin{document}

\title{\textbf{Renormalizability of the dimension two gluon operator $A^{2}$ in a
class of nonlinear covariant gauges} }
\author{\textbf{V.E.R. Lemes}\thanks{%
vitor@dft.if.uerj.br} \ , \textbf{R.F. Sobreiro}\thanks{%
sobreiro@uerj.br}{\ }\ , \and \textbf{S.P. Sorella}\thanks{
sorella@uerj.br} \ , \\
\\
\textit{UERJ, Universidade do Estado do Rio de Janeiro,} \and \textit{Rua S{%
\~a}o Francisco Xavier 524, 20550-013 Maracan{\~a},} \textit{Rio de Janeiro,
Brasil}}
\maketitle

\begin{abstract}
\noindent In this work we discuss a class of nonlinear covariant
gauges for Yang-Mills theories which enjoy the property of being
multiplicatively renormalizable to all orders. This property
follows from the validity of a linearly broken identity, known as
the ghost Ward identity. Furthermore, thanks to this identity, it
turns out that the local composite dimension two gluon operator
$A_{\mu }^{a}A_{\mu }^{a}$ can be introduced in a
mulptiplicatively renormalizable way.
\end{abstract}

\newpage

\section{Introduction}

 According to the Faddeev-Popov procedure, the
quantization of gauge theories requires a choice of the gauge fixing
to get rid of spurious degrees of freedom. This choice does not
affect physical quantities since they correspond to gauge invariant
operators. After choosing a gauge, one has to prove the
renormalizability of the theory, in order to consistently define it
at the quantum level. Moreover, in addition to the $BRST$
invariance, and depending on the choice of the gauge fixing, the
resulting theory might display further global symmetries which
reduce the number of free parameters present in the gauge fixing
term. \\\\In this letter we present a covariant nonlinear gauge
fixing which enjoys the property of being multiplicative
renormalizable, while containing a unique gauge parameter $\alpha $.
This feature stems form the existence of an additional global
symmetry amounting to perform a constant shift of the Faddeev-Popov
ghost field $c$, provided there is a compensating transformation of
the Lagrange multiplier field $b$. This additional invariance gives
rise to a linearly broken identity, known as the ghost Ward identity
\cite{Blasi:1990xz}, which ensures the renormalizability of the
theory. An interesting feature of this nonlinear gauge is that it
allows for the introduction of the local dimension two operator
$A_{\mu }^{a}A_{\mu }^{a}$, which turns out to be multiplicatively
renormalizable too. This provides an example of a nonlinear gauge
which allows for the introduction of the operator $A_{\mu
}^{a}A_{\mu }^{a}$, which has attracted much attention in recent
years. \\\\ Although in this paper the quantization of the gauge
theories is achieved through the introduction of a gauge fixing, it
is worth mentioning that the approach based on the exact
renormalization group has been proven to be a useful tool in order
to obtain a manifestly gauge invariant formulation of Yang-Mills
theories, without the need of a gauge fixing, see for instance
\cite{Arnone:2005fb,Rosten:2006qx,Morris:2006in} and references
therein.
\\\\The work is organized as follows. In section 2 we introduce
the nonlinear gauge fixing, and we derive the set of Ward
identities. In section 3 we prove the renormalizability of the
model within the framework of the algebraic renormalization
\cite{Piguet:1995er}. In section 4 we consider the inclusion of
the dimension two composite gluon operator $A_{\mu }^{a}A_{\mu
}^{a}$ and we establish its multiplicative renormalizability.
Finally, the conclusions are displayed in section 5.

\section{Gauge fixing}

The Yang-Mills action in four-dimensional Euclidean space-time is
\begin{equation}
S_{YM}=\frac{1}{4}\int d^{4}xF_{\mu \nu }^{a}F_{\mu \nu }^{a}\;,  \label{ym}
\end{equation}
with the field strength given by
\begin{equation}
F_{\mu \nu }^{a}=\partial _{\mu }A_{\nu }^{a}-\partial _{\nu }A_{\mu
}^{a}+gf^{abc}A_{\mu }^{b}A_{\nu }^{c}\;,  \label{fstr}
\end{equation}
where $g$ is the coupling constant and $f^{abc}$ are the structure constants
of $SU(N)$, the color index $a$ belongs to the adjoint representation, $%
a=1,\ldots ,N^{2}-1$.\newline
\newline
To quantize the action (\ref{ym}), we follow the $BRST$ procedure and we
introduce the Faddeev-Popov ghost and antighost fields, respectively, $c^{a}$
and $\bar{c}^{a}$, as well as the Lagrange multiplier $b^{a}$. We require
invariance of the gauge fixed action under the nilpotent $BRST$
transformations
\begin{eqnarray}
sA_{\mu }^{a} &=&-D_{\mu }^{ab}c^{b}\;,  \nonumber \\
sc^{a} &=&\frac{g}{2}f^{abc}c^{c}c^{c}\;,  \nonumber \\
s\bar{c}^{a} &=&b^{a}\;,  \nonumber \\
sb^{a} &=&0\;,  \label{brs1}
\end{eqnarray}
where the covariant derivative is defined as
\begin{equation}
D_{\mu }^{ab}=\delta ^{ab}\partial _{\mu }-gf^{abc}A_{\mu }^{c}\;.
\label{cov}
\end{equation}
We also impose that the gauge fixed action,
\begin{equation}
S=S_{YM}+S_{gf}\;,  \label{gfaction}
\end{equation}
obeys the integrated ghost equation
\begin{equation}
\mathcal{G}^{a}S=\int d^{4}x\left( \frac{\delta S}{\delta c^{a}}+gf^{abc}%
\bar{c}^{b}\frac{\delta S}{\delta b^{c}}\right) =0\;,  \label{ghost0}
\end{equation}
which expresses in a functional form the existence of an additional global
invariance, corresponding to a shift of the ghost field by a constant
together with the compensating transformation of the Lagrange multiplier $%
b^{a}$ \cite{Blasi:1990xz}. Thus, the most general gauge fixing
term compatible with both $BRST$ symmetry (\ref{brs1}) and ghost
equation (\ref{ghost0}) is found to be
\begin{eqnarray}
S_{gf} &=&s\int d^{4}x\;\bar{c}^{a}\left( \partial _{\mu }A_{\mu }^{a}+\frac{%
\alpha }{2}b^{a}+\frac{\alpha }{2}gf^{abc}\bar{c}^{b}c^{c}\right)   \nonumber
\\
&=&\int d^{4}x\left[ b^{a}\left( \partial _{\mu }A_{\mu }^{a}+\frac{\alpha }{%
2}b^{a}+\alpha {g}f^{abc}\bar{c}^{b}c^{c}\right) +\bar{c}^{a}\partial _{\mu
}D_{\mu }^{ab}c^{b}+\frac{\alpha }{4}g^{2}f^{abc}f^{cde}\bar{c}^{a}\bar{c}%
^{b}c^{d}c^{e}\right] \;.  \label{gaugefix1}
\end{eqnarray}
In order to write down the Ward identities fulfilled by the gauge
fixed action, we introduce two external sources, $\Omega _{\mu
}^{a}$ and $L^{a}$, coupled to the nonlinear $BRST$
transformations \cite{Piguet:1995er}. Thus, for the complete
starting action $\Sigma $ we get
\begin{equation}
\Sigma =S_{YM}+S_{gf}+S_{ext}\;,  \label{action2}
\end{equation}
where
\begin{eqnarray}
S_{ext} &=&s\int d^{4}x\left( -\Omega^a_{\mu }A_{\mu }^{a}+L^{a}c^{a}\right)
\nonumber \\
&=&\int d^{4}x\left( -\Omega^a_{\mu }D_{\mu }^{ab}c^{b}+\frac{g}{2}%
f^{abc}L^{a}c^{b}c^{c}\right) \;.  \label{ext}
\end{eqnarray}
\newline
As it is easily checked, the action $\Sigma $ obeys the following Ward
identities:\newline

\begin{itemize}
\item  the Slavnov-Taylor identity
\begin{equation}
\mathcal{S}(\Sigma )=\int d^{4}x\left( \frac{\delta \Sigma }{\delta \Omega
_{\mu }^{a}}\frac{\delta \Sigma }{\delta A_{\mu }^{a}}+\frac{\delta \Sigma }{%
\delta L^{a}}\frac{\delta \Sigma }{\delta c^{a}}+b^{a}\frac{\delta \Sigma }{%
\delta \bar{c}^{a}}\right) =0\;,  \label{st1}
\end{equation}

\item  the linearly broken integrated ghost Ward identity
\begin{equation}
\mathcal{G}^{a}\Sigma =\Delta _{cl}^{a}\;,  \label{ghost1}
\end{equation}
where
\begin{equation}
\Delta _{cl}^{a}=gf^{abc}\int d^{4}x\left( \Omega _{\mu }^{b}A_{\mu
}^{c}-L^{b}c^{c}\right) \;,  \label{ghost2}
\end{equation}
is a classical breaking \cite{Blasi:1990xz} linear in the fields
and $\mathcal{G}^{a}$ is given by (\ref{ghost0}).\newline For
further use, the quantum numbers of the fields and sources are
displayed in table \ref{table1}.
\begin{table}[t]
\centering
\begin{tabular}{|c|c|c|c|c|c|c|}
\hline
fields / sources & $A$ & $c$ & $\bar{c}$ & $b$ & $\Omega$ & $L$ \\ \hline
dimension & $1$ & 0 & $2$ & $2$ & $3$ & $4$ \\
ghost number & 0 & 1 & $-1$ & 0 & $-1$ & $-2$ \\ \hline
\end{tabular}
\caption{Dimension and ghost number of the fields and sources.}
\label{table1}
\end{table}
\end{itemize}

\section{Renormalizability of the model}

\noindent Let us discuss now the renormalizability of the action
(\ref {action2}). Following the framework of the algebraic
renormalization \cite{Piguet:1995er}, we shall look at the most general invariant local counterterm $%
\Sigma ^{c}$ compatible with the Ward identities characterizing
the model. From equations (\ref{st1}) and (\ref{ghost1}) it
follows thus that $\Sigma ^{c}$ fulfills the following constraints
\begin{equation}
\mathcal{B}_{\Sigma }\Sigma ^{c}=0\;,  \label{f1}
\end{equation}
\begin{equation}
\mathcal{G}^{a}\Sigma ^{c}=0\;,  \label{constcount}
\end{equation}
where $\mathcal{B}_{\Sigma }$ is the nilpotent linearized Slavnov-Taylor
operator
\begin{equation}
\mathcal{B}_{\Sigma }=\int d^{4}x\left( \frac{\delta \Sigma }{\delta \Omega
_{\mu }^{a}}\frac{\delta }{\delta A_{\mu }^{a}}+\frac{\delta \Sigma }{\delta
A_{\mu }^{a}}\frac{\delta }{\delta \Omega _{\mu }^{a}}+\frac{\delta \Sigma }{%
\delta L^{a}}\frac{\delta }{\delta c^{a}}+\frac{\delta \Sigma }{\delta c^{a}}%
\frac{\delta }{\delta L^{a}}+b^{a}\frac{\delta }{\delta \bar{c}^{a}}\right)
\;.  \label{linst1}
\end{equation}
From the general results on the cohomology of Yang-Mills theories,
see \cite{Piguet:1995er} and references therein, it follows that
the most general solution of eq.(\ref{f1}) can be written as
\begin{equation}
\Sigma ^{c}=a_{0}S_{YM}+\mathcal{B}_{\Sigma }\Delta ^{-1}\;,  \label{count1}
\end{equation}
where $\Delta ^{-1}$ is an integrated polynomial in the fields and sources,
with dimensions bounded by four and negative ghost number, namely
\begin{equation}
\Delta ^{-1}=\int d^{4}x\left( a_{1}\partial _{\mu }\bar{c}^{a}A_{\mu
}^{a}+a_{2}L^{a}c^{a}+a_{3}\frac{\alpha }{2}\bar{c}^{a}b^{a}+a_{4}\frac{%
\alpha }{2}gf^{abc}\bar{c}^{a}\bar{c}^{b}c^{c}+a_{5}\Omega _{\mu }^{a}A_{\mu
}^{a}\right) \;,  \label{triv1}
\end{equation}
where $a_{0},$ $a_{1},$ $a_{2},$ $a_{3},$ $a_{4},$ $a_{5}$ are free
coefficients. Furthermore, from the eq.(\ref{constcount}), it follows that $%
a_{4}=-2a_{3}$ and $a_{2}=0$. Thus, the most general invariant counterterm
turns out to contains four free parameters, $a_{0},$ $a_{1},$ $a_{3},$ $a_{5}
$, being given by
\begin{equation}
\Sigma ^{c}=a_{0}S_{YM}+\mathcal{B}_{\Sigma }\int d^{4}x\left[ a_{1}\partial
_{\mu }\bar{c}^{a}A_{\mu }^{a}+a_{3}\frac{\alpha }{2}\bar{c}^{a}\left(
b^{a}-2gf^{abc}\bar{c}^{b}c^{c}\right) +a_{5}\Omega _{\mu }^{a}A_{\mu
}^{a}\right] \;.  \label{mcfct}
\end{equation}
After having characterized the most general counterterm, it remains to check
the stability of the action (\ref{action2}), amounting to prove that the
counterterm $\Sigma ^{c}$ can be reabsorbed by a multiplicative redefinition
of the parameters, fields, and sources of $\Sigma $, according to
\begin{equation}
\Sigma (\Phi ,J,\xi )+\epsilon \Sigma ^{c}(\Phi ,J,\xi )\;=\Sigma (\Phi
_{0},J_{0},\xi _{0})+O(\epsilon )\;,  \label{ren}
\end{equation}
with
\begin{equation}
\begin{array}{ccc}
\Phi _{0} & = & Z_{\Phi }^{1/2}\Phi \;, \\
J_{0} & = & Z_{J}J\;, \\
\xi _{0} & = & Z_{\xi }\xi \;,
\end{array}
\;\;\;\;
\begin{array}{ccc}
\Phi  & \in  & \left\{ A,c,\bar{c}\right\} \;, \\
J & \in  & \left\{ \Omega ,L\right\} \;, \\
\xi  & \in  & \left\{ g,\alpha \right\} \;,
\end{array}
\label{ren1}
\end{equation}
In fact, by direct inspection one finds
\begin{eqnarray}
Z_{A}^{1/2} &=&1+\epsilon \left( \frac{a_{0}}{2}+a_{5}\right) \;,  \nonumber
\\
Z_{c}^{1/2}=Z_{\bar{c}}^{1/2} &=&1-\epsilon \frac{a_{1}}{2}\;,  \nonumber \\
Z_{g} &=&1-\epsilon \frac{a_{0}}{2}\;,  \nonumber \\
Z_{\alpha } &=&1+\epsilon \left( a_{0}+2a_{1}+a_{3}\right) \;,  \label{ren3}
\end{eqnarray}
and
\begin{eqnarray}
Z_{b}^{1/2} &=&1-\epsilon \left( \frac{a_{0}}{2}+a_{1}\right)
\;=\;Z_{g}Z_{c}\;,  \nonumber \\
Z_{\Omega } &=&Z_{g}^{-1}Z_{A}^{-1/2}Z_{c}^{-1/2}\;,  \nonumber \\
Z_{L} &=&Z_{g}^{-1}Z_{c}^{-1}\;,  \label{ren4}
\end{eqnarray}
thus establishing the multiplicative renormalizability of the
action $\Sigma $.

\section{Inclusion of the dimension two gluon operator}

\noindent Let us discuss now the inclusion of the dimension two
gluon operator $A_{\mu }^{a}A_{\mu }^{a}$ in the case of the
nonlinear gauge (\ref {gaugefix1}). According to
\cite{Verschelde:2001ia,Browne:2003uv,Dudal:2002pq}, we add the
operator $A_{\mu }^{a}A_{\mu }^{a}$ to the action (\ref{action2})
through the following term
\begin{eqnarray}
S_{LCO} &=&s\int d^{4}x\left( \lambda \frac{A_{\mu }^{a}A_{\mu }^{a}}{2}+%
\frac{\zeta }{2}\lambda {J}\right) \;,  \nonumber \\
&=&\int d^{4}x\left( J\frac{A_{\mu }^{a}A_{\mu }^{a}}{2}+\lambda {A}_{\mu
}^{a}\partial _{\mu }{c}^{a}+\frac{\zeta }{2}J^{2}\right) \;,  \label{lco}
\end{eqnarray}
where $\lambda $ and $J$ are external sources introduced as a BRST doublet
\begin{eqnarray}
s\lambda  &=&J\;,  \nonumber \\
sJ &=&0\;.  \label{brs2}
\end{eqnarray}
The quantum numbers of the external sources are displayed in table \ref
{table2}.
\begin{table}[t]
\centering
\begin{tabular}{|c|c|c|}
\hline
LCO sources & $\lambda$ & $J$ \\ \hline
dimension & 2 & 2 \\
ghost number & $-1$ & 0 \\ \hline
\end{tabular}
\caption{Dimension and ghost number of the LCO sources.}
\label{table2}
\end{table}
The dimensionless parameter $\zeta $ is needed in order to take
into account the ultraviolet divergences present in the Green
function $\left\langle A^{2}(x)A^{2}(y)\right\rangle $
\cite{Verschelde:2001ia,Browne:2003uv,Dudal:2002pq}. The action we
will work with  is now given by
\begin{equation}
\Sigma ^{\prime }=\Sigma +S_{LCO}\;,  \label{action3}
\end{equation}
where $\Sigma $ stands for expression (\ref{action2}). The
introduction of the operator $A_{\mu }^{a}A_{\mu }^{a}$ does not
affect the $BRST$ symmetry and the ghost equation.  In fact, the
modified action $\Sigma ^{\prime }$ obeys the following
Slavnov-Taylor identity
\begin{equation}
\mathcal{S}(\Sigma ^{\prime })=\int d^{4}x\left( \frac{\delta \Sigma
^{\prime }}{\delta \Omega _{\mu }^{a}}\frac{\delta \Sigma ^{\prime }}{\delta
A_{\mu }^{a}}+\frac{\delta \Sigma ^{\prime }}{\delta L^{a}}\frac{\delta
\Sigma ^{\prime }}{\delta c^{a}}+b^{a}\frac{\delta \Sigma ^{\prime }}{\delta
\bar{c}^{a}}+J\frac{\delta \Sigma ^{\prime }}{\delta \lambda }\right) =0\;,
\label{st2}
\end{equation}
while the ghost equation (\ref{ghost1}) remains unaffected,
\begin{equation}
\mathcal{G}^{a}\Sigma ^{\prime }=\Delta _{cl}^{a}\;,  \label{ghost3}
\end{equation}
with $\mathcal{G}^{a}$ and $\Delta _{cl}^{a}$ given, respectively,
by (\ref {ghost0}) and (\ref{ghost2}).\newline
\newline
The multiplicative renormalizability of the action (\ref{action3}) can be
established in the same way as that of the action (\ref{action2}). For that,
one looks at the most general invariant counterterm $\widetilde{{\Sigma }}$,
which is an integrated polynomial in the fields and sources with dimension
bounded by four and with vanishing ghost number. From the Slavnov-Taylor
identity (\ref{st2}), it follows that $\widetilde{{\Sigma }}$ can be written
as
\begin{equation}
\widetilde{{\Sigma }}=a_{0}S_{YM}+\mathcal{B}_{\Sigma }^{\prime }\Delta
^{-1}\;,  \label{count2}
\end{equation}
where $\Delta ^{-1}$ reads
\begin{eqnarray}
\Delta ^{-1} &=&\int d^{4}x\left( a_{1}\partial _{\mu }\bar{c}^{a}{A}_{\mu
}^{a}+a_{2}L^{a}c^{a}+a_{3}\frac{\alpha }{2}\bar{c}^{a}b^{a}+a_{4}\frac{%
\alpha }{2}gf^{abc}\bar{c}^{a}\bar{c}^{b}c^{c}+a_{5}\Omega _{\mu }^{a}A_{\mu
}^{a}\right.   \nonumber \\
&+&\left. a_{6}\frac{\lambda }{2}A_{\mu }^{a}A_{\mu }^{a}+a_{7}\frac{\zeta }{%
2}\lambda {J}+a_{8}\bar{c}^{a}c^{a}\right) \;,  \label{triv2}
\end{eqnarray}
and $\mathcal{B}_{\Sigma }^{\prime }$ is the linearized nilpotent operator
corresponding to the Slavnov-Taylor identity (\ref{st2}), namely
\begin{equation}
\mathcal{B}_{\Sigma }^{\prime }=\int d^{4}x\left( \frac{\delta \Sigma
^{\prime }}{\delta \Omega _{\mu }^{a}}\frac{\delta }{\delta A_{\mu }^{a}}+%
\frac{\delta \Sigma ^{\prime }}{\delta A_{\mu }^{a}}\frac{\delta }{\delta
\Omega _{\mu }^{a}}+\frac{\delta \Sigma ^{\prime }}{\delta L^{a}}\frac{%
\delta }{\delta c^{a}}+\frac{\delta \Sigma ^{\prime }}{\delta c^{a}}\frac{%
\delta }{\delta L^{a}}+b^{a}\frac{\delta }{\delta \bar{c}^{a}}+J\frac{\delta
}{\delta \lambda }\right) \;.  \label{linst2}
\end{equation}
Moreover, the ghost identity (\ref{ghost3}) implies that $\widetilde{{\Sigma
}}$ is constrained by
\begin{equation}
\mathcal{G}^{a}\widetilde{{\Sigma }}=0\;,  \label{ghost4}
\end{equation}
from which it follows that $a_{2}=0$, $a_{4}=-2a_{3}$ and $a_{8}=0$. Thus,
for the final form of the counterterm we get
\begin{eqnarray}
\widetilde{\Sigma } &=&a_{0}S_{YM}+\mathcal{B}_{\Sigma }^{\prime }\int
d^{4}x\left[ a_{1}\partial _{\mu }\bar{c}^{a}{A}_{\mu }^{a}+a_{3}\frac{%
\alpha }{2}\bar{c}^{a}\left( b^{a}-2gf^{abc}\bar{c}^{b}c^{c}\right)
+a_{5}\Omega _{\mu }^{a}A_{\mu }^{a}\right.   \nonumber \\
&+&\left. a_{6}\frac{\lambda }{2}A_{\mu }^{a}A_{\mu }^{a}+a_{7}\frac{\zeta }{%
2}\lambda J\right] \;.  \label{count3}
\end{eqnarray}
As done in the previous section, we have to check that the counterterm $%
\widetilde{\Sigma }$ corresponds to a redefinition of the fields,
sources and parameters of the action $\Sigma ^{\prime }$. In fact,
it turns out that the action (\ref{action3}) is multiplicatively
renormalizable. The renormalization of the fields, $BRST$ sources
and coupling constant are given
as in eqs.(\ref{ren1}), (\ref{ren3}) and (\ref{ren4}). Also, the parameter $%
\zeta $ and the sources $\lambda ,J$ renormalize as
\begin{eqnarray}
\zeta _{0} &=&Z_{\zeta }\zeta \;,  \nonumber \\
\lambda _{0} &=&Z_{\lambda }\lambda \;,  \nonumber \\
J_{0} &=&Z_{J}J\;.  \label{ren5}
\end{eqnarray}
with
\begin{eqnarray}
Z_{\zeta } &=&1+\epsilon \left( 2a_{0}-2a_{6}+a_{7}\right) \;,  \nonumber \\
Z_{\lambda } &=&1+\epsilon \left( -\frac{a_{0}}{2}+\frac{a_{1}}{2}%
+a_{6}\right) \;=\;Z_{g}^{-1}Z_{c}^{-1}Z_{J}\;,  \nonumber \\
Z_{J} &=&1+\epsilon \left( a_{6}-a_{0}\right) \;.  \label{ren6}
\end{eqnarray}
Notice, in particular, that the source $J$, and thus the composite operator $%
A_{\mu }^{a}A_{\mu }^{a}$ coupled to it, displays multiplicative
renormalizability.

\section{Discussion and conclusions}

\noindent In this work we have discussed a class of nonlinear
covariant gauges characterized by the validity of the integrated
broken ghost Ward identity \cite{Blasi:1990xz}. This identity,
together with the Slavnov-Taylor identity, has enabled us to prove
the multiplicative renormalizability of the theory, a feature which
has been established to all orders of perturbation theory by means
of the algenraic renormalization \cite{Piguet:1995er}. Further, we
have been able to introduce the dimension two gluon operator $A_{\mu
}^{a}A_{\mu }^{a}$, while maintaining the renormalizability of the
model. \\\\  The example of the covariant linear gauges presented
here enlarges the number of gauges for which a local dimension two
operator can be introduced in a multiplicatively renormalizable way.
Such a dimension two operator can be, in fact, introduced in many
gauges, namely: the Landau gauge
\cite{Verschelde:2001ia,Browne:2003uv,Dudal:2002pq}, the linear
covariant gauges \cite{Dudal:2003by}, the Curci-Ferrari gauge
\cite{Dudal:2003gu}, the maximal Abelian gauge\footnote{In the case
of the Curci-Ferrari and maximal Abelian gauges a generalized
dimension two operator, {\it i.e.} $\left( \frac{A^{2}}{2}+\alpha
\overline{c}c\right) $, has to be considered.}
\cite{Kondo:2001tm,Dudal:2004rx}, as well as in a variety of
interpolating gauges \cite{Dudal:2004rx,Capri:2005zj}.
\\\\Although the operator $A_{\mu }^{a}A_{\mu }^{a}$ is not gauge
invariant, the property of being multiplicatively renormalizable
in a rather large number of gauges can be interpreted as evidence
in favor of its relevance for the infrared behavior of the gluon
propagator. It is worth reminding here that the condensate
$\left\langle A_{\mu }^{a}A_{\mu }^{a}\right\rangle $ is in fact
directly related to the appearance of an effective dynamical gluon
mass, a topic which is receiving increasing attention in recent
years
\cite{Lavelle:1988eg,Gubarev:2000nz,Gubarev:2000eu,Verschelde:2001ia,Kondo:2001nq,
Gracey:2004bk,Li:2004te,Boucaud:2001st,Boucaud:2002nc,Boucaud:2005rm,RuizArriola:2004en}.
\\\\ Finally, let us mention that the issue of the quantization of
Yang-Mills theories in nonlinear gauges has also been investigated
from other viewpoints. For example, in \cite{Alkofer:2003jr} one
finds a detailed analysis of the nonlinear Curci-Ferrari gauge
within the framework of the Schwinger-Dyson equations, where the
gluon propagator has been found to the suppressed in the infrared.
Recently, a potential application of these gauges on the lattice
has been advocated in \cite{Ghiotti:2006pm}.

\section*{Acknowledgments}

The Conselho Nacional de Desenvolvimento Cient\'{i}fico e Tecnol\'{o}gico
(CNPq-Brazil), the Faperj, Funda{\c{c}}{\~{a}}o de Amparo {\`{a}} Pesquisa
do Estado do Rio de Janeiro, the SR2-UERJ and the Coordena{\c{c}}{\~{a}}o de
Aperfei{\c{c}}oamento de Pessoal de N{\'{i}}vel Superior (CAPES) are
gratefully acknowledged for financial support.

\end{document}